\documentclass[runningheads]{llncs}
\usepackage[T1]{fontenc}
\usepackage{amsfonts}
\usepackage{graphicx,verbatim}
\usepackage{amssymb}
\usepackage{subcaption}

\usepackage{booktabs}
\usepackage{amsmath}
\begin{document}
%

\title{Multi-scale radiomics in pelvic MRI for endometriosis subtyping: highlighting data heterogeneity constraints}

\titlerunning{MRI Radiomics for Endometriosis Subtyping}
%
\author{Eliot Leguy\inst{1,2} \and
Chloe Mallet\inst{2} \and
Nicolas Normand\inst{2} \and Elodie Germani\inst{1}}
\authorrunning{Leguy, Mallet et al.}
%
\institute{Laboratoire Traitement du Signal et de l’Image (LTSI, INSERM UMR 1099), Université de Rennes, Rennes, France \and
Nantes Université, Ecole Centrale Nantes, CNRS, LS2N, UMR 6004, Nantes, France \\
Corresponding author: \email{elodie.germani@univ-rennes.fr}}

\maketitle              

\begin{abstract}
Analyzing female pelvic MRIs is challenging, especially for evaluating endometriosis, where visual features are influenced by several factors, including anatomical complexity, technical variability, and inter-reader variability. Here, we evaluate a radiomics-based pipeline for patient-level endometriosis subtyping using the publicly available UT-EndoMRI dataset. We extract radiomics features from manually segmented uterine and ovarian regions and compare several multi-scale feature representations and feature-selection strategies. We train supervised classifiers to distinguish patients with at least one endometrioma from those without, and perform an unsupervised perturbation analysis to assess whether radiomics profiles reveal reproducible patient subgroups. The best supervised performance is achieved using raw Wavelet-derived features and a Gradient Boosting classifier, yielding an AUC of 0.80. However, this model produces several false positives, resulting in low specificity. ComBat harmonization does not consistently improve performance, suggesting that post hoc harmonization is insufficient in a small, multi-site cohort in which acquisition groups contained very few patients. Using an unsupervised clustering analysis, we identify reproducible but poorly separated partitions that remain associated with acquisition variables. Overall, these results suggest that pelvic MRI radiomics contain a preliminary signal for endometriosis subtyping, while highlighting the fragility of radiomics-based subtyping in small, multi-site datasets.
\end{abstract}

\section{Introduction}

Endometriosis is a chronic inflammatory disease affecting around 10\% of women worldwide and characterized by the presence of endometrial-like tissue (also called ``lesions'') outside the uterus. Disease diagnosis and treatment planning require detailed evaluation of lesions, including their location, extension, and tissue composition~\cite{chapron2019rethinking}. To this end, pelvic magnetic resonance imaging (MRI) is a central tool for a non-invasive evaluation of endometriosis~\cite{bazot2004deep}. Recently, scoring systems such as the deep pelvic endometriosis index (dPEI)~\cite{thomassin2020magnetic} have helped standardize this evaluation, providing a reproducible pipeline for characterizing endometriosis status and facilitating treatment planning. However, this scoring system relies on visual examination of pelvic MRI to locate lesions and measure their extent. However, due to the complexity of the female pelvic anatomy, the impact of acquisition parameters on image appearance, and the level of expertise required to identify and locate endometriosis lesions, this process remains challenging and prone to inter-reader variability, especially for junior readers~\cite{MICELI2026112926}.

In this context, radiomics can offer a complementary quantitative approach to visual examination by extracting structured image descriptors, including shape, intensity, and texture features, from regions of interest~\cite{vanGriethuysen2017ComputationalRadiomics}. Multi-scale filtering, such as Wavelet decomposition and Laplacian of Gaussian (LoG) filtering, can further characterize spatial patterns at different frequencies. In female pelvic imaging, radiomics-based methods have shown promising results for assessing adenomyosis~\cite{burla2024mri}, classifying endometrial lesions~\cite{liu2023development}, and differentiating ovarian endometriomas from ovarian dermoid cysts~\cite{liu2024ultrasound}. 

However, radiomics features are sensitive to several sources of variability, such as technical variability, in which scanner differences, acquisition protocols, reconstruction algorithms, and image resolution can alter feature distributions independently of pathology. In multi-site studies, harmonization methods such as ComBat~\cite{orlhac2022combat} have been proposed to reduce technical variability while preserving clinically relevant covariates. In addition, standard radiomics feature-extraction pipelines produce a large number of non-independent features that must be filtered for subsequent use. Feature selection methods such as LASSO~\cite{tibshirani1996lasso} and Boruta~\cite{kursa2010boruta} can help limit the effects of high dimensionality in small cohorts and evaluate residual technically related bias.

Here, we evaluate the potential and challenges of radiomics for endometriosis subtyping from a publicly available multi-centric pelvic MRI dataset~\cite{liang_utendomri_2025}. Our contributions are threefold. First, we develop a pipeline for multi-scale radiomics feature extraction from organ segmentations in T1w and T2w pelvic MRIs and train machine-learning classifiers to distinguish patients with and without endometrioma. Second, we explore the effects of inter-site harmonization on performance using ComBat in settings with small sample sizes per site and unknown covariate effects. Third, we perform an unsupervised analysis of radiomics perturbation profiles to investigate whether the feature space reveals clinically meaningful subgroups or is primarily driven by technical variability. 

\section{Methods}

\subsection{Radiomics feature extraction pipeline}
Let $\mathcal{I}_{i,s}$ denote the MRI three-dimensional (3D) volume of patient $i$ for sequence $s$, and let $\Omega_{i,r} \subset \mathbb{Z}^3$ denote the segmented region of interest (ROI) corresponding to anatomical region $r$. Radiomics extraction is defined as the mapping
\[
\Phi: (\mathcal{I}_{i,s}, \Omega_{i,r}) \rightarrow \mathbf{x}_{i,s,r} \in \mathbb{R}^{p},
\]
where $\mathbf{x}_{i,s,r}$ is a vector of image-derived descriptors.

To assess the contribution of multi-scale information, radiomics features are extracted from the original and filtered images. Wavelet filtering decomposes each image into eight 3D sub-bands $(LLL, LLH, \dots, HHH)$, where $L$ and $H$ denote low-pass and high-pass filtering along each spatial axis. These decompositions capture complementary low-frequency anatomical structure and high-frequency local texture. LoG filtering is used to emphasize local intensity transitions at multiple spatial scales. For an image $\mathcal{I}$, the LoG response at scale $\sigma$ is defined as
\(
L_{\sigma} = \nabla^2(G_{\sigma} \circledast \mathcal{I}),
\)
where $G_{\sigma}$ is a Gaussian kernel.

\subsection{Inter-site harmonization}

To reduce the effects of technical variability on radiomics features, we explore the benefits of ComBat harmonization~\cite{orlhac2022combat}. For each radiomics feature, ComBat models the observed value as:
\[
y_{ij} = \alpha + \mathbf{x}_{ij}^{\top}\boldsymbol{\beta} + \gamma_i + \delta_i \varepsilon_{ij},
\]
where \(y_{ij}\) is the value of the feature measured for subject or region \(j\) in acquisition setting \(i\), \(\alpha\) is the global feature mean, \(\mathbf{x}_{ij}^{\top}\boldsymbol{\beta}\) represents the effect of preserved covariates, \(\gamma_i\) is an additive batch effect, \(\delta_i\) is a multiplicative batch effect, and \(\varepsilon_{ij}\) is the residual error. In our case, \(\mathbf{x}_{ij}^{\top}\boldsymbol{\beta}\) represents the expected contribution of the endometrioma label to the radiomics feature value, so that this disease-related information is preserved while correcting for acquisition-related effects.

After estimating the batch parameters, the corrected feature value are:
\[
y^{\mathrm{ComBat}}_{ij}
=
\frac{y_{ij} - \hat{\alpha} - \mathbf{x}_{ij}^{\top}\hat{\boldsymbol{\beta}} - \hat{\gamma}_i}{\hat{\delta}_i}
+
\hat{\alpha}
+
\mathbf{x}_{ij}^{\top}\hat{\boldsymbol{\beta}},
\]
so that site-related additive and multiplicative effects are reduced while the specified covariate effects are preserved. Although ComBat is widely used for imaging biomarker harmonization, it depends on several assumptions. Reliable estimation of site effects requires sufficient samples per acquisition site and balanced covariate distributions. Previous work~\cite{orlhac2022combat,jodoin2025combat} has recommended between 20 and 30 subjects per site, while more recent studies suggest that reasonable harmonization can be obtained with 16 to 32 subjects in a moving site when a well-populated reference site is available. ComBat also assumes that covariate effects are comparable across sites. Violations of this assumption, or strong confounding between site and disease, can lead to erroneous harmonization or removal of clinically relevant signals. Here, we analyze ComBat-harmonized features alongside raw features rather than using them as a guaranteed correction.

\subsection{Feature selection}
Radiomics features extraction produces a high-dimensional feature space with substantial redundancy between descriptors. To address these challenges, we perform a two-step feature selection strategy. 
First, we use LASSO logistic regression to identify sparse linear predictors, retaining only features with non-zero coefficients. Second, we use Boruta~\cite{kursa2010boruta} to identify features with non-linear predictive relevance. This method trains a Random Forest classifier on the original features augmented with randomized shadow features. A feature is retained only if its importance is consistently higher than the importance of randomized shadow features. For each configuration, the final signature is defined as the intersection of the LASSO-selected and Boruta-selected features. 

\subsection{Supervised classification}\label{sec:supervised_classification}
The supervised task is formulated as a binary classification problem. For each patient-level observation, the label is defined as
\(y_i \in \{0,1\},\)
where $y_i=1$ denotes a patient with at least one confirmed endometrioma and $y_i=0$ denotes a patient without any identified endometrioma. The objective is to evaluate whether radiomics features extracted from pelvic MRI can discriminate endometrioma-positive from endometrioma-negative patients. We evaluate four candidate classifiers: Logistic Regression, Random Forest Classifier with 100 trees, Support Vector Machine with radial basis function kernel, and Gradient Boosting classifier. Each model is embedded in a feature standardization pipeline. Candidate models are compared using five-fold cross-validation on the training set, with the area under the receiver operating characteristic curve (AUC) as the criterion for selecting the final model. The selected model is retrained on the whole training set and evaluated on the held-out test set. Evaluation metrics include AUC, accuracy, sensitivity, specificity, precision, and F1-score.

\subsection{Unsupervised Perturbation Analysis}
To further assess the discriminative power of radiomics features, we perform an unsupervised evaluation of patient-level radiomics profiles. We define the reference control population as a sample of 8 patients, identified from the cohort data as having neither endometriosis nor endometrioma. Because the dataset reports the number of patients without endometriosis but does not explicitly provide their identifiers, this control group was derived from the available metadata and should be interpreted cautiously. For each feature, patient values were converted into robust perturbation scores relative to this control population:
\[
z_{i,j}
=
\frac{x_{i,j} - \mathrm{median}(X_{\mathrm{control},j})}
{\mathrm{IQR}(X_{\mathrm{control},j})},
\]
where $X_{\mathrm{control},j}$ denotes the distribution of feature $j$ among the inferred control patients. Features with zero interquartile range in the control group are excluded.

We then perform dimensionality reduction using Principal Component Analysis (PCA), retaining the smallest number of components explaining 90\% of the total variance. We apply K-means clustering to these components for $k \in \{2, \dots, 6\}$, and select the optimal number of clusters by maximizing the silhouette coefficient. Cluster stability is assessed using 50 bootstrap samples and the Adjusted Rand Index (ARI). To evaluate whether the resulting clusters reflect clinical structure or residual acquisition bias, associations between cluster labels and technical variables, such as imaging site and scanner model, are quantified using $\chi^2$ tests and Cramer's $V$.

\section{Experiments}

\begin{figure}[t]
\begin{subfigure}[b]{0.48\textwidth}
\centering
\scriptsize
\begin{tabular}{|l|c|}
\hline
\textbf{Characteristic} & \textbf{Value} \\
\hline
Total patients & 51 \\
Endometrioma-positive patients & 40 \\
Endometrioma-negative patients & 11 \\
Primary MRI sequences & T2, T1FS \\
Analyzed anatomical regions & Ovary, uterus \\
\hline
\end{tabular}
\caption{Dataset characteristics.} \label{tab:dataset_summary}
\end{subfigure}
\begin{subfigure}[b]{0.48\textwidth}
\centering
\includegraphics[width=0.48\textwidth]{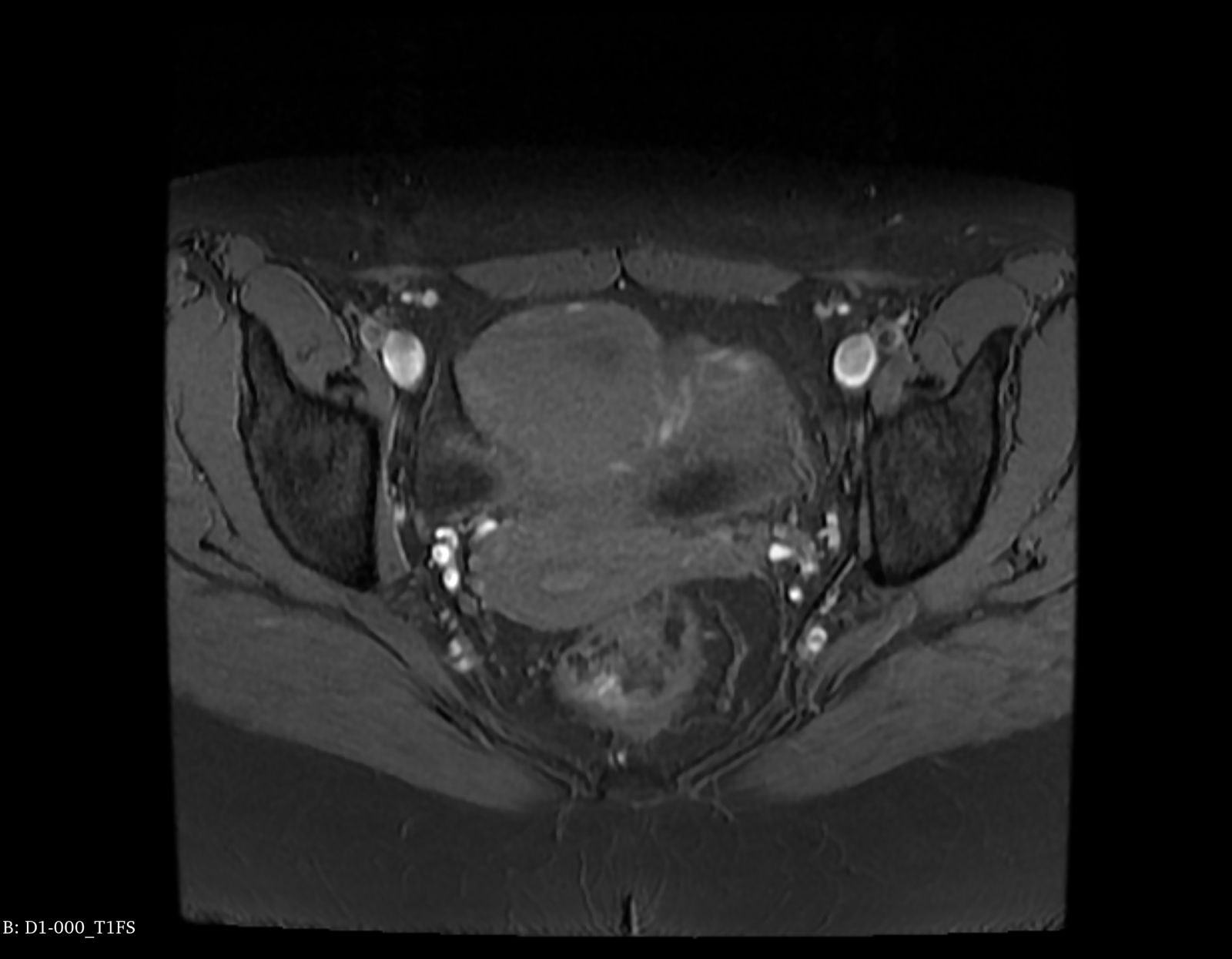} \includegraphics[width=0.48\textwidth]{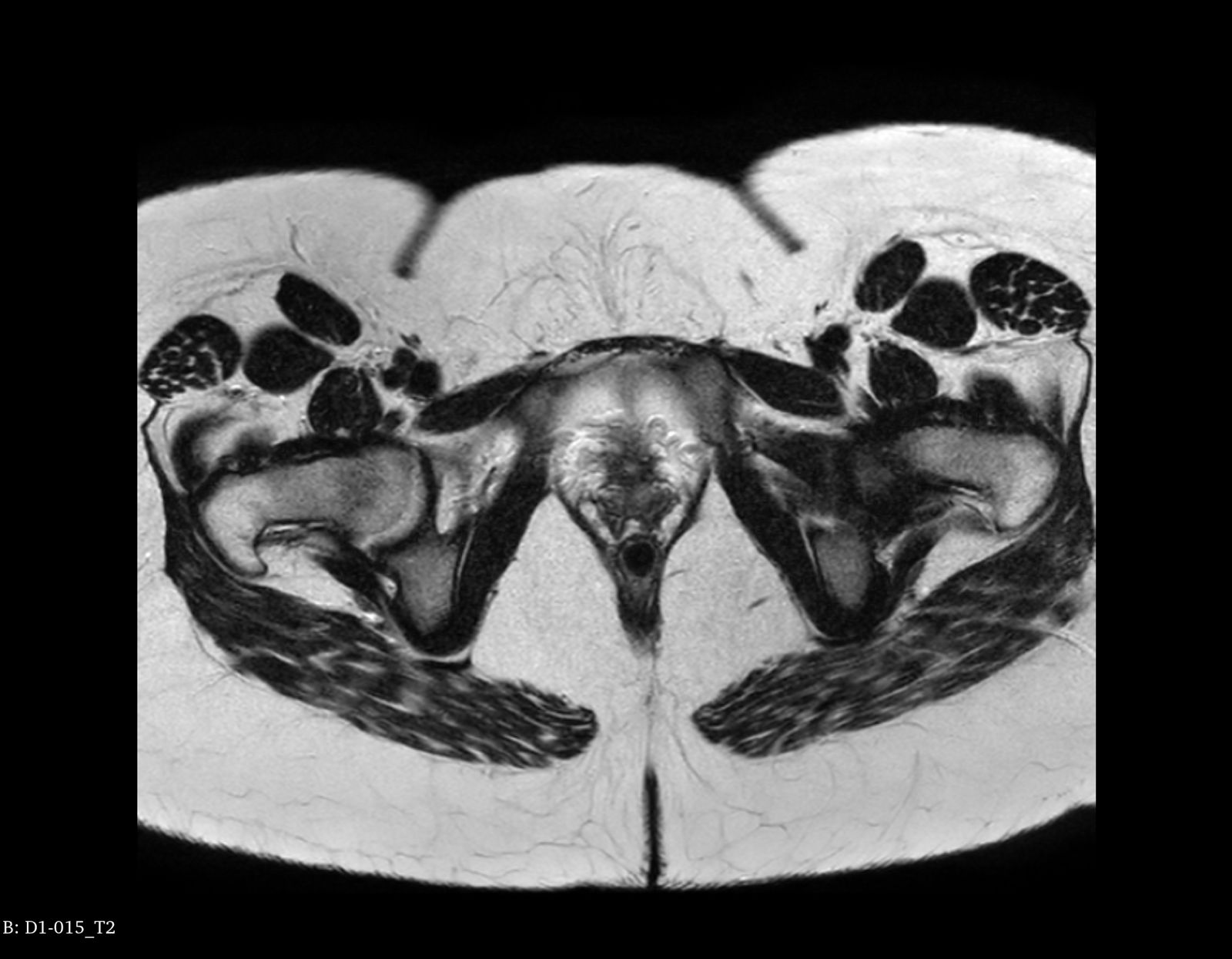}
\caption{Left: T1 FatSat, Right: T2 slices.}
\label{fig:ex-mri}
\end{subfigure}
\caption{Summary of the UT-EndoMRI D1 cohort.}
\end{figure}

\paragraph{\textbf{Dataset}}
We conduct our experiments on the UTHealth Endometriosis MRI dataset (UT-EndoMRI)~\cite{liang_utendomri_2025}, a publicly available dataset containing pelvic multi-sequence MRI and manual organs and lesions segmentations from women with endometriosis. The complete dataset is organized into two cohorts. Here, we use the first cohort, denoted D1, which contains 51 patients as described in Tab~\ref{tab:dataset_summary}, acquired before 2022 from the Memorial Hermann Hospital System and the Texas Children's Hospital Pavilion for Women. We use T2- and T1-weighted fat-suppressed (T1FS) MRI sequences and manual segmentations of the uterus and ovaries. Radiomics features from both organs are aggregated at the patient level. This dataset comprises 51 patients, of whom 40 are endometrioma-positive, and 11 are endometrioma-negative. We construct a balanced held-out test set containing 5 positive and 5 negative patients. The remaining patients (N=41, 35 positive and 6 negative) are used for training and cross-validation.

Tab.~\ref{tab:dataset_summary} summarizes the dataset used for the classification experiments.


\paragraph{\textbf{Implementation details}}
In the supervised classification task, we compare models trained with four radiomics configurations: Wavelet, Wavelet+ComBat, LoG, and LoG+ComBat. Features are extracted using PyRadiomics v3.0.1~\cite{vanGriethuysen2017ComputationalRadiomics}, following the Image Biomarker Standardization Initiative (IBSI)~\cite{ref_ibsi_url}. The extracted features include shapes, first-order statistics, GLCM, GLRLM, GLSZM, and GLDM descriptors. MRI volumes are normalized to a scale of 100 and resampled to an isotropic $1.0$~mm spacing using B-spline interpolation. All preprocessing and feature extraction are performed in the volume space. In the LoG configuration, Gaussian scales are set to $\sigma \in \{1.0, 2.0, 3.0\}$~mm. When multiple segmentation masks are available for the same patient, organ, and sequence, feature values are averaged to produce a single patient-level representation for each configuration. Missing or non-finite values are handled by mean imputation fitted on the full training set, followed by training-fitted z-score standardization. To prevent leakage, ComBat parameters are estimated on the training set only and then applied to the held-out test set. No test-set information is used to estimate these harmonization parameters. Feature selection is performed on the full train set once and kept fixed for the rest of the experiments. For classifier selection, we use cross-validation on this fixed training-derived set of features, while final performance is assessed on the held-out test set. Feature selection is not repeated inside each cross-validation fold. Therefore, cross-validation is used to optimize the classifier on a fixed training-derived signature, while final generalization is assessed only on the independent held-out test set.

\subsection{Supervised Classification Results}
Tab.~\ref{tab:classification_results} summarizes the performance of the four radiomics configurations. The highest cross-validation AUC is obtained with ComBat-harmonized Wavelet features and Logistic Regression (AUC=$0.955 \pm 0.065$). However, the highest held-out test AUC is obtained with raw Wavelet features and Gradient Boosting (AUC=0.80). With this configuration, the model detects positive cases, but its low specificity also indicates many false positives. Given the 10-patient test set, the difference between raw and ComBat-harmonized Wavelet features should not be overinterpreted. These results do not show a clear benefit of ComBat in this cohort, possibly because site- and scanner-specific effects are estimated from small and imbalanced acquisition groups. LoG-based models perform worse than Wavelet-based models, with test AUCs of 0.52 without harmonization and 0.58 after ComBat. Overall, these results suggest preliminary radiomics signals for endometriosis subtyping in Wavelet-derived texture features of uterine and ovarian masks. However, the discrepancy between cross-validation and held-out performance shows that the models' performance remains unstable. Given the small size of the test set, these results should be interpreted as exploratory rather than as evidence of clinical generalization.

\begin{table}[t]
\caption{Performance of radiomics classifiers for endometriosis subtyping. CV AUC is reported as mean $\pm$ standard deviation across cross-validation folds. Other metrics are computed on the held-out test set.}
\label{tab:classification_results}
\centering
\begin{tabular}{|l|c|c|c|c|c|c|c|c|}
\hline
\textbf{Configuration} & \textbf{Model} & \textbf{CV AUC} & \textbf{Test AUC} & \textbf{Acc.} & \textbf{Sens.} & \textbf{Spec.} & \textbf{Prec.} & $\mathbf{F_1}$ \\
\hline
Wavelet raw & GB & $0.936 \pm 0.108$ & 0.800 & 0.60 & 1.000 & 0.200 & 0.556 & 0.714 \\
Wavelet ComBat & LR & $0.955 \pm 0.065$ & 0.680 & 0.60 & 0.800 & 0.400 & 0.571 & 0.667 \\
LoG raw & LR & $0.864 \pm 0.135$ & 0.520 & 0.50 & 0.400 & 0.600 & 0.500 & 0.444 \\
LoG ComBat & RF & $0.929 \pm 0.101$ & 0.580 & 0.60 & 0.800 & 0.400 & 0.571 & 0.667 \\
\hline
\end{tabular}
\end{table}

\subsection{Radiomic feature analysis}
The selected features include shape, first-order, and texture descriptors, suggesting that the classification signal is driven not only by global morphology but also by intensity distribution and local spatial heterogeneity. In the Wavelet configurations, selected descriptors are mainly derived from high-frequency sub-bands, indicating that directional multi-scale texture information contributes to the supervised signal observed in the classification experiments. LoG-derived signatures retained features are related to local intensity transitions, but the corresponding models show weaker performance. 

\begin{figure}[t]
    \centering \scriptsize
    Raw features  \hspace{3cm} ComBat harmonized features \\ \includegraphics[width=0.5\textwidth]{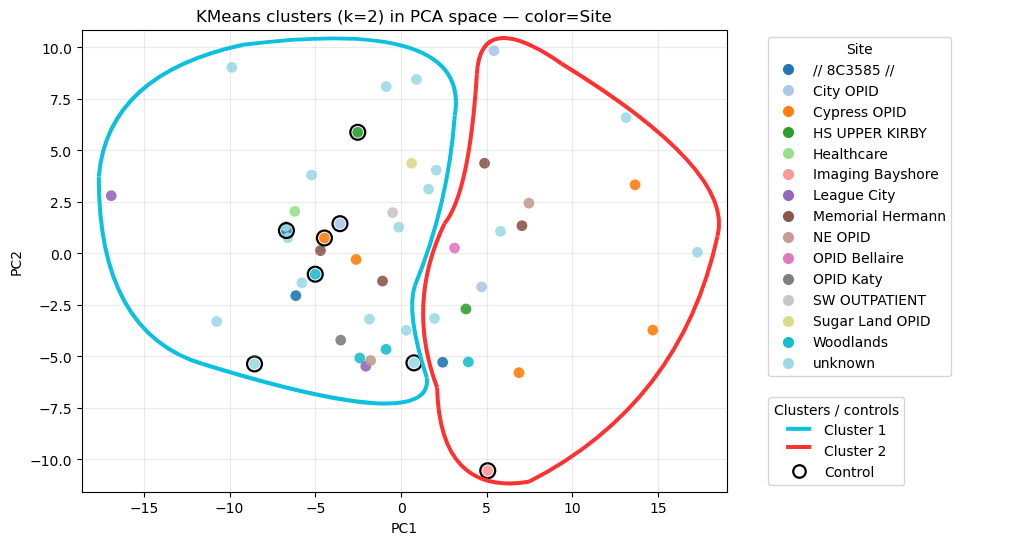}\hfill \includegraphics[width=0.5\textwidth]{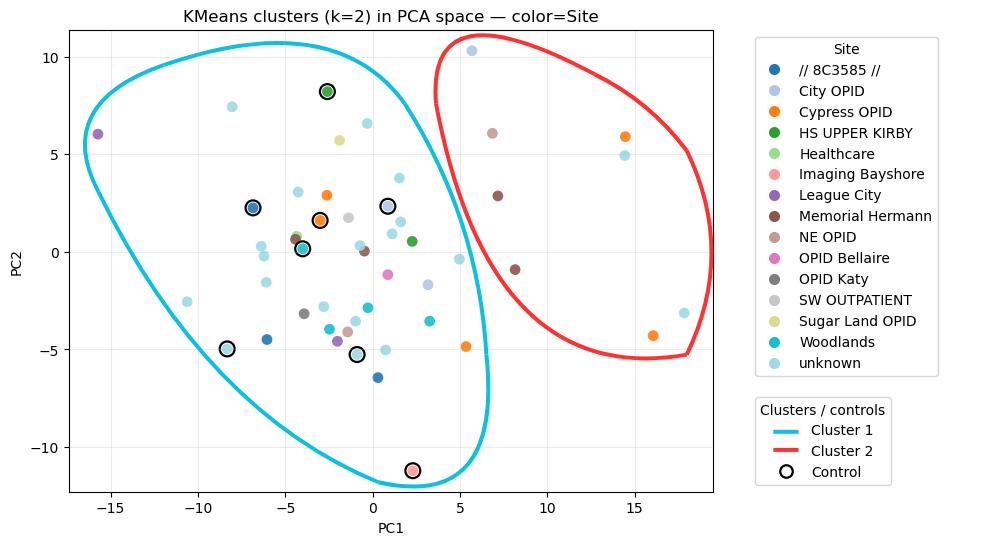}
    \caption{Projection of the first two components from the PCA of radiomics perturbation profiles after K-means clustering with $k=2$. The control patients (circled dots) have no endometriomas. Dots' colors represent their acquisition site.}
    \label{fig:pca_clustering}
\end{figure}

\subsection{Unsupervised clustering of radiomics perturbation profiles}
To investigate whether the radiomics feature space contains clinically relevant patient subgroups, we perform an unsupervised clustering analysis based on perturbation profiles. Each patient is represented by a z-score vector computed relative to a control population without endometriosis or endometrioma. 

\begin{table}[t]
\caption{Quality and technical association of the unsupervised clustering results: best silhouette score, Bootstrap stability measured by ARI, and the association with acquisition site using $\chi^2$ $p$-values and Cramer's $V$.}
\label{tab:clustering_quality}
\centering 
\begin{tabular}{|l|c|c|c|c|c|}
\hline
\textbf{Configuration} & $\mathbf{k}$ & \textbf{Silh.} & \textbf{ARI K-means} & \textbf{Site $(p;V)$} & \textbf{Scanner $(p;V)$} \\
\hline
Raw, all organs & 2 & 0.371 & $0.859 \pm 0.095$ & $0.461;\,0.569$ & $0.386;\,0.417$ \\
ComBat, all organs & 2 & 0.318 & $0.900 \pm 0.087$ & $0.106;\,0.616$ & $0.310;\,0.417$ \\
Raw, ovary & 2 & 0.241 & $0.699 \pm 0.154$ & $0.200;\,0.652$ & $0.186;\,0.560$ \\
Raw, uterus & 2 & 0.278 & $0.806 \pm 0.177$ & $0.118;\,0.649$ & $0.665;\,0.467$ \\
ComBat, ovary & 2 & 0.330 & $0.879 \pm 0.168$ & $\mathbf{0.022};\,0.768$ & $0.528;\,0.480$ \\
ComBat, uterus & 2 & 0.263 & $0.769 \pm 0.276$ & $0.412;\,0.529$ & $0.164;\,0.680$ \\
\hline
\end{tabular}
\end{table}

The clustering analysis does not seem to reveal distinct clinical phenotypes. Across all configurations, the silhouette criterion is optimal for $k=2$, but the corresponding silhouette scores remain modest, ranging from $0.241$ to $0.371$. Since clearly separated clusters are typically expected to have substantially higher silhouette scores, often above $0.7$, these results suggest weak separation and should not be interpreted as evidence of robust clinical subtypes. The PCA projections in Fig.~\ref{fig:pca_clustering} show substantial overlap between the two groups. Together with the modest silhouette scores reported in Tab.~\ref{tab:clustering_quality}, this suggests that the dominant radiomic structure corresponds to a broad perturbation gradient rather than to well-separated patient subtypes. Despite this limited separation, the two-cluster partitions are relatively stable under resampling. Bootstrap ARI values for K-means range from $0.699 \pm 0.154$ to $0.900 \pm 0.087$, indicating moderate to high clustering stability. This combination of modest silhouette scores and stable ARI suggests that the feature space contains a reproducible partition, but not one that should automatically be interpreted as clinically meaningful. The association analysis further supports this cautious interpretation. Several configurations show high Cramer's $V$ values for the acquisition variables, even when the corresponding $\chi^2$ tests are not statistically significant, likely due to the limited cohort size. The strongest site association is observed after harmonization in the ovary-specific analysis ($p=0.022$, $V=0.768$), indicating that ComBat does not fully remove technical structure from the radiomics space. Therefore, the clustering results suggest a residual acquisition signature rather than evidence of reproducible endometriosis subtypes.

\section{Conclusion}
This study evaluates a radiomics-based pipeline for patient-level endometriosis subtyping from pelvic MRI. We compare multi-scale radiomics features based on Wavelet and LoG filtering, with and without ComBat harmonization, and obtain the best held-out performance with raw Wavelet features, reaching a test AUC of 0.80 on a balanced test set, suggesting that Wavelet-derived texture features might contain discriminative information for endometriosis subtyping.

However, cross-validation performance is consistently higher than held-out performance, suggesting model instability in the small-sample setting. Inter-site harmonization with ComBat does not improve generalization: it reduces the test AUC for Wavelet features and yields only a modest improvement for LoG features. This observation is likely related to the limited number of patients per acquisition site or scanner group, often ranging from 1 to 10, which is below the commonly recommended sample sizes for reliable ComBat estimation. 
Clustering of radiomics perturbation profiles does not reveal clearly separated clinical phenotypes. Instead, clusters are strongly associated with acquisition variables such as site and scanner, indicating that a substantial part of the radiomics structure reflects technical variability rather than disease-specific subtypes.

Overall, this work provides preliminary evidence that multi-scale radiomics features, particularly Wavelet-derived descriptors, can contribute to MRI-based endometriosis subtyping. At the same time, it highlights the fragility of radiomics models in small multi-center cohorts and the practical limits of post hoc harmonization when site-level sample sizes are small. Future work should focus on larger external validation cohorts, stricter control of acquisition protocols, better-balanced site distributions, and more robust domain-adaptation or harmonization strategies before such models can be considered clinically reliable.

\newpage

%
%
%
\bibliographystyle{splncs04}
\bibliography{rapport}
%




\end{document}